\documentclass[twocolumn,pre,showpacs]{revtex4}
\usepackage{amsmath,amssymb,bm,epsfig}

\begin{document}
\title{Jamming and pattern formation in models of segregation}
\author{Tim Rogers and Alan J. McKane}\affiliation{Theoretical Physics Division, School of Physics \& Astronomy, The University of Manchester, Manchester M13 9PL, UK}
\begin{abstract}
We investigate the Schelling model of social segregation, formulated as an intrinsically non-equilibrium system, in which the agents occupy districts (or patches) rather than sites on a grid. We show that this allows the equations governing the dynamical behaviour of the model to be derived. Analysis of these equations reveals a jamming transition in the regime of low-vacancy density, and inclusion of a spatial dimension in the model leads to a pattern forming instability. Both of these phenomena exhibit unusual characteristics which may be studied through our approach.
\end{abstract}
\maketitle
\section{Introduction}
Forty years ago Thomas Schelling proposed a simple model of social segregation in which agents of two different types are placed at random on a grid, before being allowed to move according to a desire to be close to other agents of the same type \cite{Schelling1969}. This seminal work  played an important part in the development of the new scientific field of social simulation, in which sociological problems are studied through computational models.\par
To the physics community, there is an immediate similarity between the rules of the Schelling model and some of the simple dynamical models of statistical physics and the tools of this field have been used to provide quantitative insight into the behavior of the model \cite{DallAsta2008}. There have also been several attempts to make direct links to equilibrium spin models \cite{Stauffer2007,Grauwin2009,Gauvin2009,Gauvin2010}, though care must be taken as agents in the Schelling model are subject to kinetic constraints which mean that the dynamics cannot be viewed as a simple energy minimisation process \cite{DallAsta2008}. In what follows we will show that certain analytically tractable Schelling-like models of segregation exhibit a range of interesting physical phenomena.\par
In Schelling's original work \cite{Schelling1969}, and most subsequent studies \footnote{There are a wealth of variations of Schelling model, many of which fit into the unified framework outlined in \cite{Rogers2011ii}; we refer to that paper for a broader review of the literature.}, the city in which the agents reside is modeled as a 2-dimensional grid. We propose instead to study a model in which the basic object of interest is a district (patch) containing multiple residences. This modification improves the social realism of the model -- a city is much better described as a collection of districts and suburbs with their own ethnic character than as a simple grid. Moreover, patch models are naturally amenable to analytical approaches, for example in ecology \cite{McKane2004}. Indeed, a patch variation of a Schelling-like model has been considered before \cite{Grauwin2009}, though that work has a very different flavor to our own, being primarily concerned with applying the techniques equilibrium statistical mechanics. We choose the dynamical rules of our model to be close to those of the lattice-based model of Gauvin \textit{et al} \cite{Gauvin2009}, simulations of which display interesting physical behavior that we intend to study theoretically.\par
Our analysis is divided into two parts, investigating different implementations of the patch-based Schelling model. In both cases we consider a large city, divided into $N$ patches each containing $K$ residences. For the first model we investigate, model A, the patches are relatively small, but there are very many of them. Through enumerating the possible interactions between patches, we obtain a description of the model in terms of a deterministic dynamical system. This framework is used to investigate a jamming transition present in the model, in which large numbers of agents remain stuck in unfavorable states. We then go on to analyse a model with a spatial structure, model B, and consider the alternative limit of very large patches. The behavior in this case is rather different, with the model exhibiting pattern formation driven by antidiffusion. 

\section{Model A}

We begin by considering the situation where $K$ is relatively small; each patch represents a local neighborhood containing only a few residences. Initially, the city is randomly populated with equal numbers of agents of two different types, which we call $A$ and $B$, with a fraction $\rho$ of the residences left vacant. At each time step two residences are chosen at random from the whole city. If the first contains an agent and the second is vacant, then that agent is given the opportunity to move to the vacant residence. They take up this offer only if the number of agents of the opposite type in the destination patch is less than a threshold $T$.

The contents of patch $i$ at time $t$ is encoded in the numbers $a_i(t)$, $b_i(t)$ and $v_i(t)$ of $A$ agents, $B$ agents and vacancies it contains. The state of the system as a whole is then specified (up to trivial reordering of the residences) by the quantities
\begin{equation}
F_{a,b,v}(t)=\frac{1}{N}\sum_{i=1}^N \delta_{a,a_i(t)}\delta_{b,b_i(t)}\delta_{v,v_i(t)}\,,
\end{equation}
giving the fraction of patches in state $(a,b,v)$ at time $t$. Our theoretical work is based on an analysis of the time evolution of these quantities when the number of patches is very large. 

The first step is to consider the possible changes to a given $F_{a,b,v}$ which can occur in one time step. Suppose, for example, that at time $t$ an $A$ agent in a patch with state $(a,b,v)$ is selected to move to a vacancy in a patch with state $(a',b',v')$. This event occurs with probability $$F_{a,b,v}(t)\frac{a}{K}F_{a',b',v'}(t)\frac{v'}{K}\Theta(T-b')\,. $$ The factors in this expression are explained as (i) the probability of choosing a patch in state $(a,b,v)$, (ii) the probability of selecting an $A$ agent from that patch, (iii) the probability of choosing a patch in state $(a',b',v')$ for the destination, (iv) the probability of selecting a vacant residence in the destination patch, and lastly (v) a Heaviside $\Theta$ function imposing the requirement that the destination patch contains fewer than $T$ agents of type $B$. As a result of this interaction, the values of $F_{a,b,v}(t)$ and $F_{a',b',v'}(t)$ would decrease by $1/N$, whilst $F_{a-1,b,v+1}(t)$ and $F_{a'+1,b,v'-1}(t)$ would increase by $1/N$. The effects of moving a $B$ agent are computed in the same way. 

In the present model, the patches are well-mixed in the sense that each patch is equally likely to interact with every other; it follows that when the number of patches is very large, it is sufficient to consider only the average over all possible interactions. Rescaling time by a factor of $1/N$, we take the thermodynamic limit $N\to\infty$, in which the random quantities $F_{a,b,v}(t)$ may be well approximated by continuous deterministic functions of rescaled time. In this formalism the possible changes that can occur in one time step discussed above are translated into a deterministic differential equation which exactly describes the behavior of the system in the limit $N\to\infty$. Summing over the possible interactions gives
\begin{equation}
\frac{d}{dt}F_{a,b,v}(t)=\sum_{a',b',v'}\hspace{-3pt} F_{a',b',v'}(t)\,\Big(R_A^++R_B^++R_A^-+R_B^-\Big)\,,
\label{Pdot}
\end{equation}
where the contributions from each type of interaction are
\begin{figure}
\includegraphics[width=240pt, trim=20 15 20 0]{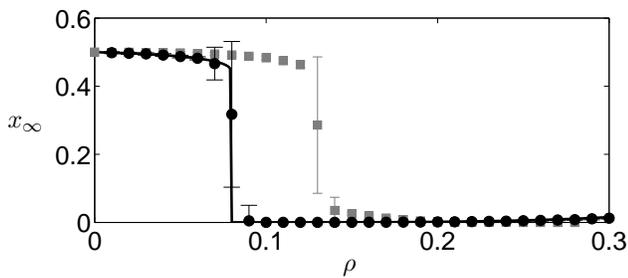}
\caption{\label{Fig1}Equilibrium interface density $x_\infty$ as a function of vacancy density $\rho$, for patches of size $K=9$ and tolerance parameter $T=3$. Theoretical prediction (solid line) from equation (\ref{Pdot}) is compared to results from stochastic simulations on $N=10^3$ patches (black circles) and on the lattice (grey squares), averaged over 100 samples. Error bars are one standard deviation.}
\end{figure}
\begin{equation*}
\begin{array}{lcr}
R_A^+&=&\displaystyle\bigg. \frac{a'}{K}\Big(\mathcal{E}^-_a\mathcal{E}^+_v-1\Big)\Big[F_{a,b,v}(t)\frac{v}{K}\Theta(T-b)\Big]\,\,\,\\
R_B^+&=&\displaystyle\bigg. \frac{b'}{K}\Big(\mathcal{E}^-_b\mathcal{E}^+_v-1\Big)\Big[F_{a,b,v}(t)\frac{v}{K}\Theta(T-a)\Big]\,\,\,\\
R_A^-&=&\displaystyle\bigg. \frac{v'}{K}\Theta(T-b')\Big(\mathcal{E}^+_a\mathcal{E}^-_v-1\Big)\Big[F_{a,b,v}(t)\frac{a}{K}\Big]\,\,\,\\
R_B^-&=&\displaystyle\bigg. \frac{v'}{K}\Theta(T-a')\Big(\mathcal{E}^+_b\mathcal{E}^-_v-1\Big)\Big[F_{a,b,v}(t)\frac{b}{K}\Big]\,.\\
\end{array}
\end{equation*}
The $\mathcal{E}^\pm$ used here are step operators which alter the functions they act on through the addition or subtraction of 1 to their argument; for example, $\mathcal{E}^-_a\mathcal{E}^+_v\big[F_{a,b,v}(t)\big]=F_{a-1,b,v+1}(t)$.

To monitor the emergence of segregation in the model, we measure the fraction of pairs of neighbouring agents of different types, a statistic commonly referred to as the interface density \cite{DallAsta2008,Gauvin2010,Rogers2011ii}. A patch in state $(a,b,v)$ has $(a+b)(a+b-1)/2$ distinct pairs of agents, of which $ab$ are $(a,b)$ pairs. The interface density $x$ is found by summing over all patches, or alternatively by the formula 
\begin{equation*}
x(t)=\sum_{a,b,v}F_{a,b,v}(t)\frac{2ab}{(a+b)(a+b-1)}\,.
\end{equation*}
In simulations of lattice-based versions of the model (without patches), behavior suggesting a phase transition in interface density has been observed as the vacancy density $\rho$ is lowered; below a critical value, there are not enough vacancies to facilitate the movement of agents to a segregated state and the system appears `jammed' \cite{Edmonds2005,Gauvin2009}. We can confirm the existence of this transition in the patch model under investigation here by numerically integrating (\ref{Pdot}) to find the final state of the system. 

Starting from a well-mixed initial condition for the $F_{a,b,v}$ (chosen to be equivalent to the large $N$ limit of a random initial configuration of agents in the microscopic model) we numerically integrate (\ref{Pdot}) using the forward-Euler scheme. Figure 1 shows the equilibrium value $x_\infty$ of interface density as a function of $\rho$ for patches of size $K=9$ and a tolerance parameter $T=3$. 

For low values of vacancy density the segregated steady state becomes inaccessible to the dynamics started from a well-mixed initial condition, and the system finds a different (well-mixed) equilibrium; we estimate the critical point for the transition to be $\rho_c\approx0.079615$. Further information about the system can be gained through a linear stability analysis of the steady states reached from both well-mixed and segregated initial conditions. In the jammed regime we find two stable equilibria, one well-mixed and one segregated, moreover, the stability of both states increases with $\rho$. Past the transition point only a single, segregated, steady state can be found. 

In figure 1, the analytical result is compared with results from simulations of the patch model with $N=10^3$ patches, averaged over 100 samples. Also shown are the results from simulations of a lattice based version of the model \cite{Gauvin2009}, where the jamming transition occurs at a different point. We should point out that the transition is not unique to the values of $K$ and $T$ we have chosen; further numerical results suggest that the conditions $K\geq 3$ and $T<K/2$ are sufficient.

\begin{figure}
\includegraphics[width=240pt, trim=30 30 30 0]{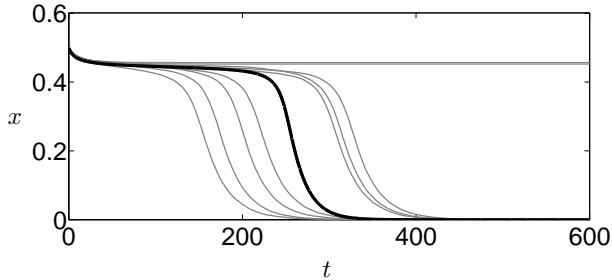}
\caption{\label{Fig2} Comparison between the deterministic theory (black line) and the results of ten simulation runs (gray lines) for vacancy density $\rho=0.08$, just above the point of the jamming transition. Patches of size $K=9$ and a tolerance parameter of $T=3$ were used for both data sets. The stochastic simulations were performed on a system of $N=10^3$ patches.}
\end{figure}
Beyond confirming the existence of the transition, we are also able to probe the behavior of the system near the critical point. Figure \ref{Fig2} shows the deterministic dynamics for a value of vacancy density just above critical, $\rho=0.08$. Also shown in that figure are several sample results from stochastic simulations of a system of size $N=10^3$. The data from simulations show very little noise, although both the final outcome (either jammed or unjammed), and the moment in time that unjamming takes place, appears random. Ordinarily, one might expect that the persistence of a metastable state followed by sudden relaxation to equilibrium is a stochastic effect which would not be captured by a naive deterministic theory. In the present case, however, we see that the deterministic theory does indeed display the same behavior, with the effect of stochasticity and finite size mainly limited to fluctuations in timing. These effects are reduced in larger system sizes.

There is an analogy \cite{Gauvin2009,Gauvin2010} between the lattice based version of the segregation model and kinetically constrained models which are used as simplified proxies in the study of glasses and granular media \cite{Garrahan2010}. Kinetically constrained models have been intensively studied over the last decade and they exhibit a rich phenomenology, including a jamming transition \cite{Toninelli2006,Biroli2008,Shokef2010} which is both discontinuous in its order parameter and features exponentially diverging relaxation time. 

\begin{figure}
\includegraphics[width=180pt, trim=10 30 20 0]{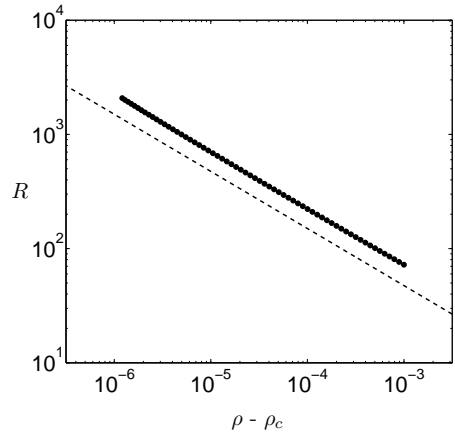}
\caption{\label{Fig3}Log-log plot of the duration of the metastable state, measured by $R$ given in equation (\ref{defR}), against $\rho-\rho_c$. Circles show the results of numerical integration of (\ref{Pdot}), the dashed line is a power-law with exponent $-1/2$.}
\end{figure}

In the present model it is also the case that, as vacancy density is further lowered towards the critical value, the waiting time until the system is freed from the metastable jammed state increases. In analogy with critical slowing of magnetization in the Ising model \cite{Schneider1972}, we introduce the following relaxation time for interface density:
\begin{equation}
R=\int_0^\infty \big(x(t)-x_\infty\big)\,dt\,,
\label{defR}
\end{equation}
viewed as a function of $\rho-\rho_c$. As shown in Figure \ref{Fig3}, this quantity does not grow exponentially as in some kinetically constrained models, but rather exhibits power-law behavior with exponent $-1/2$. 

\section{Model B}

We now introduce a simple modification to the above model which will allow us to develop a theoretical description of the emergence of spatial patterns of agents. We take the $N$ patches considered earlier and arrange them in a 1-dimensional lattice; from now on, agents are restricted to move only to patches which neighbor them in the lattice. The other rules of the model remain the same, though to aid the analysis we smooth the step function, introducing $\Theta_\kappa(x)=(1+\tanh(2x/\kappa))/2$, for a small parameter $\kappa$. 

For this spatial version of the model we employ a different analysis in which we keep the number of patches finite and take the limit of large patch size $K\to\infty$. In a sociological context this model represents a city divided into several districts, each still containing a large number of residences. A similar model featuring agents with no preferences was considered in \cite{Fanelli2010} and the same techniques apply here with the simple addition of the $\Theta_\kappa$ threshold term. 

The state of the system at a given time is specified by the vectors $\bm{a}=(a_1,\ldots a_N)$ and $\bm{b}=(b_1,\ldots b_N)$ giving the numbers of $A$ and $B$ agents in each patch. Note that there is no need to keep track of vacancies as $a_i+b_i+v_i=K$ for all $i$. The dynamics of the model are determined by the transition probabilities $P(\bm{a},\bm{b}\big|\bm{a}',\bm{b}')$, giving the likelihood of moving from state $(\bm{a}',\bm{b}')$ to state $(\bm{a},\bm{b})$ in one time step. Changes to the system result from the movement of agents between neighboring patches; the transition probabilities for $A$ and $B$ agents have the forms  
\begin{equation}
\begin{split}
&P\big(a_i-1,a_j+1\big|a_i,a_j\big)=\delta_{|i-j|,1}\frac{a_i}{NK}\frac{v_j}{2K}\Theta_\kappa(b_i-T)\,,\\
&P\big(b_i-1,b_j+1\big|b_i,b_j\big)=\delta_{|i-j|,1}\frac{b_i}{NK}\frac{v_j}{2K}\Theta_\kappa(a_i-T)\,,\\
\end{split}
\end{equation}
where $i$ and $j$ are neighbors in the lattice, and we have listed as arguments only those entries of $\bm{a}$ and $\bm{b}$ which change. Writing $\pi_t(\bm{a},\bm{b})$ for the probability of finding the system in state $(\bm{a},\bm{b})$ at time $t$, we have the equation
\begin{equation*}
\begin{split}
&\pi_{t+1}(\bm{a},\bm{b})-\pi_t(\bm{a},\bm{b})\\
&=\sum_{\bm{a}',\bm{b}'}\Big(P\big(\bm{a},\bm{b}\big|\bm{a}',\bm{b}'\big)\pi_t(\bm{a}',\bm{b}')-P\big(\bm{a}',\bm{b}'\big|\bm{a},\bm{b}\big)\pi_t(\bm{a},\bm{b})\Big)\,.
\end{split}
\end{equation*}
Multiplying by $a_i$ and summing over all states, we obtain a difference equation for the average number of $A$ agents in patch $i$:
\begin{equation*}
\begin{split}
&\langle a_i \rangle_{t+1}-\langle a_i \rangle_{t}=\\
&\sum_{j\in i}\Big\langle P\big(a_i+1,a_j-1\big|a_i,a_j\big)- P\big(a_i-1,a_j+1\big|a_i,a_j\big)\Big\rangle_t\,,
\end{split}
\end{equation*}
where the notation $j\in i$ indicates that the sum is over all patches $j$ which are nearest neighbors of patch $i$. A similar expression exists for $B$ agents.  We now rescale time by a factor of $1/K$ and introduce
\begin{equation*}
\alpha_i=\frac{\langle a_i \rangle }{K}\,,\quad \beta_i=\frac{\langle b_i \rangle }{K}\,,\quad \gamma_i=\frac{\langle v_i \rangle }{K}\,,\quad\textrm{and}\quad \tau=\frac{T}{K}\,.
\end{equation*}
Taking the limit $K\to\infty$ transforms the difference equations for $\langle a_i \rangle_{t}$ and $\langle b_i \rangle_{t}$ into a pair of coupled reaction-diffusion equations:
\begin{equation}
\begin{array}{lcr}
\dot{\alpha}_i&=&\gamma_i\Theta_\kappa(\tau-\beta_i)\Delta \alpha_i - \alpha_i\Delta\big[\bigg.\gamma_i\Theta_\kappa(\tau-\beta_i)\big]\,,\\
\dot{\beta}_i&=&\gamma_i\Theta_\kappa(\tau-\alpha_i)\Delta \beta_i - \beta_i\Delta\big[\gamma_i\Theta_\kappa(\tau-\alpha_i)\big]\,,\\
\end{array}
\label{dotab}
\end{equation}
where $\Delta$ is the discrete Laplacian.

By taking the limit of a large number of agents per patch, the emergence of segregation in this model is reduced to the question of stability of the homogeneous state $\alpha_i=\beta_i=(1-\rho)/2$ for all $i$. Diagonalizing the Jacobian of (\ref{dotab}) at this point, we find that there is once again there is a jamming transition: for fixed $\tau<1/2$ and $\kappa\ll1$, the homogeneous state is stable for small $\rho$, becoming unstable as $\rho$ approaches $1-2\tau$.

\begin{figure}
\includegraphics[width=240pt, trim=0 10 0 -10]{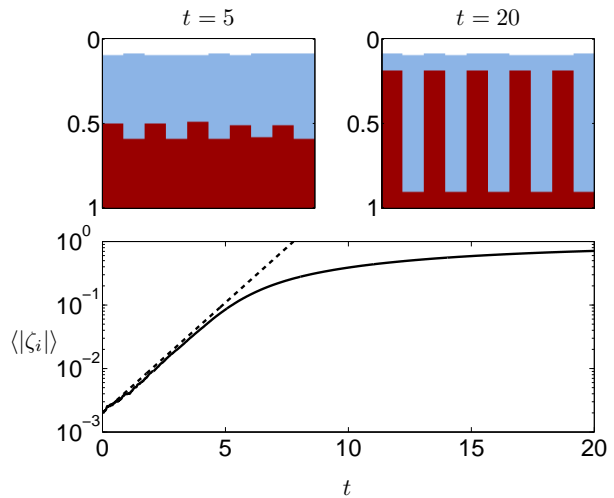}
\caption{\label{Fig4}(Color online) Upper -- two snapshots taken from a simulation of the spatial patch model on a 1-dimensional toroidal lattice of $N=10$ patches, with patch size $K=10^6$, vacancy density $\rho=0.1$ and step parameter $\kappa=0.1$. The fraction of agents of type $A$ and $B$ in each patch are shown by darker (red) and lighter (blue) areas, respectively, with vacancy density in white. Lower -- time evolution of the average absolute difference in agent densities $\langle|\zeta_i|\rangle=\frac{1}{N}\sum_i|\alpha_i-\beta_i|$ (solid line), alongside the linearised result (dashed line) from the theory.}
\end{figure}

In numerical simulations starting with a uniform distribution of agents in the un-jammed phase, the model rapidly forms a distinct pattern of alternate patches filled by agents of different types. The emergence of these patterns can be studied theoretically by setting $\rho=1-2\tau$ and analysing the stability of a homogeneous distribution of agents (that is, $\alpha_i=\beta_i=(1-\rho)/2$ for all $i$). 

In the analysis of typical pattern forming systems, it is normal to study the continuum version of the reaction-diffusion equations \cite{Murray2002}. The homogeneous state may be found to be unstable with respect to perturbations which are periodic in space; the eventual shape of the pattern which forms is then given by the most unstable wavelength. In the present model we find a different behavior, with instability on the scale of lattice spacing making it inappropriate to take the continuum limit.

The linearised form of (\ref{dotab}) can be decoupled by considering vacancy densities $\gamma_i$ and a conjugate variable $\zeta_i=\alpha_i-\beta_i$. Near the homogeneous fixed point we find
\begin{equation}
\begin{split}
&\dot{\gamma}_i=\frac{1}{2}\left(1+\frac{\rho(1-\rho)}{\kappa}\right)\Delta \gamma_i\,,\\
&\dot{\zeta}_i=\frac{1}{2}\left(\rho-\frac{\rho(1-\rho)}{\kappa}\right)\Delta\zeta_i\,.\\
\end{split}
\label{decoupled}
\end{equation}

From the first line above we infer that, near the homogeneous state, the vacancies will exhibit diffusion, independent of the behavior of the agents. For the conjugate variables $\zeta_i$ the behavior is opposite: note that the coefficient in the second line is typically large and \emph{negative}, indicating rapid anti-diffusion.
By taking an initial condition in the form of an alternating perturbation $\alpha_i=(1-\rho)/2+\varepsilon(-1)^i$ and $\beta_i=(1-\rho)/2+\varepsilon(-1)^{i+1}$ for a small positive $\varepsilon$, we find that the lattice dependence drops out, giving 
\begin{equation*}
\gamma_i(t)= \rho\,,\quad\textrm{and}\quad \zeta_i(t)=2\varepsilon(-1)^i e^{-\left(\rho-\frac{\rho(1-\rho)}{\kappa}\right)\,t}\,.
\end{equation*}

Together these facts give a description of the emergence of an alternating pattern of patches dominated by agents of different types, which develop without altering the distribution of vacancies. Initially the pattern will grow exponentially quickly before being limited by the effect of patches becoming saturated with one type of agent, at which time the stochastic model will deviate from the results of the anti-diffusive linearised system. This picture is confirmed by comparison to numerical simulations, as shown in Figure \ref{Fig4}. 

\section{Discussion}

The models introduced and analysed above are simple coarse-grained modifications of the well-known lattice based segregation models introduced by Schelling. We have demonstrated that the well-mixed patch model reproduces the jamming transition observed in simulations of lattice models, and moreover that it is described by deterministic equations which are exact in the thermodynamic limit. One might expect that the unusual character of the jamming transition is intrinsically linked to the effects of stochasticity or spatial dimension. Our analysis shows that this is not the case; the behavior is captured by the deterministic dynamical system (\ref{Pdot}). 

It should be noted that, despite the similarities in the nature of the models and of the characteristics of the transition, the jamming found here is possibly a different phenomenon to that occurring in kinetically constrained models. Specifically, in those models jamming often refers to the (probabilistic) existence of an extensive fraction of frozen spins \cite{Toninelli2006,Biroli2008,Shokef2010}. The analysis we have presented links the occurrence of jamming in a stochastic model to a dynamical transition in a particular deterministic system. As such, we cannot draw conclusions about the behavior of individual agents in the thermodynamic limit. Another possible analogy is with quasi-stationary states in models with long range interactions \cite{Dauxois2002,Antoniazzi2007}. Whatever its precise nature though, the jamming transitions in models of segregation are certainly interesting phenomena which are worthy of further investigation.

Introducing a spatial dimension to the model and studying the limit of large patch size, we observe pattern formation driven by anti-diffusion. Once again, the model is simple enough that the theoretical analysis provides a very complete picture of the mechanisms behind the emergence of segregation.

These results exemplify the possibilities for theoretical physics analyses of social models. Whilst our emphasis has mainly been on the novel physics encountered in such systems, the predictive power of the theoretical approach we use should also be of interest to simulators. 

\section{Acknowledgements}
Thanks to Matteo Marsili, Tobias Galla and Peter Sollich for useful discussions and suggestions. This work was funded by the EPSRC under grant number EP/H02171X/1.

\end{document}